# *i* Vector used in Speaker Identification by Dimension Compactness

Soumen Kanrar
Department of Computer Science
Vidyasagar University
West Bengal, India
rscs_soumen@mail.vidyasagar.ac.in

## ABSTRACT
The automatic speaker identification procedure is used to extract features that help to identify the components of the acoustic signal by discarding all the other stuff like background noise, emotion, hesitation, etc. The acoustic signal is generated by a human that is filtered by the shape of the vocal tract, including tongue, teeth, etc. The shape of the vocal tract determines and produced, what signal comes out in real time. The analytically develops shape of the vocal tract, which exhibits envelop for the short time power spectrum. The ASR needs efficient way of extracting features from the acoustic signal that is used effectively to makes the shape of the individual vocal tract. To identify any acoustic signal in the large collection of acoustic signal i.e. corpora, it needs dimension compactness of total variability space by using the GMM mean supervector. This work presents the efficient way to implement dimension compactness in total variability space and using cosine distance scoring to predict a fast output score for small size utterance.

## Categories and Subject Descriptors
H 1.1 [**Information System**]: Models and Principles – *System and Information Theory*

## 1. INTRODUCTION

In current technology of an automatic speech recognition (ASR) system produce good prediction in controlled environment i.e. the collected samples of utterance from the clean environment. The most implementable areas of ASR system are more noise full environments such as target advertising, forensic science, and service customization. The accuracy among the previously implemented ASR system based on pure GMM (Gaussian mixture model) or HMM (Hidden Markov model) [9, 10, 11, 12, 15] brings poor performance, particularly in noise full environment. Another major drawback in old ASR procedure consumes large computation time. Number of methods has been proposed over the last decade to give an efficient technique to solve this issue in ASR system or even in the other Avenue like multimedia streaming [16, 17], but still remain it is challenging one. The previous method used EM (expectation maximization) module that consumes a lot of computation time during the final prediction about the unknown utterance. The current start of art, dimension compactness successfully reduces the computation time and efficiently works in noise full environments. In old ASR system also used the score normalization for the predicted score, based on the $\log(\text{likelihood})$ ratio test [8]. The proposed work used the cosine kernel to predict the closeness among the test utterance and target utterance that gives very fast and efficient prediction in compare to old ASR system. Najim Dehak et.al., proposed new modeling about the low dimensional speaker and channels dependent space [1]. Deep Neural Networks technique being proposed by P.Kenny et.al.,[2] for extracting statistics for speaker recognition. The environmental sound classifications are based on acoustic feature extraction been proposed by Takumi [3]. The delta spectral cepstral coefficient proposed by kshitiz kumar et .al., for robust speech recognition, [4]. Maximum likelihood estimates of the supervector covariance matrix that effectively extended speaker adaption for Eigen voice estimation [5]. The accent recognition by $i-\text{vector}$ based on Gaussian means super-vector improved the performance of ASR system [6]. The cosine based on distance scoring is currently proposed to improve the computation time for predict the existence of utterance in a collected sample [7]. This paper is structured as follows. Section I introduces about the problem. Section II presents the spectral analysis. Section III presents the dimension compactness of total variability space. Section IV presents the cosine based distance scoring between the test utterance and target utterance in the corpora of acoustic signal with the conclusion at the end.

## 2. SPECTRAL ANALYSIS

Speech waves are band limited to 4kHz, 8kHz and 16kHz respectively and sampled at 8kHz to 32kHz and windowed by the Hamming window of 20ms long with the 10ms shift.

As the audio signal is constantly changing, so it is required to simplification, in this regards we expect that on short time, audio signal doesn't change statistically. Here we can assume. It is statistically stationary, but obviously the samples are constantly changing on even in the short time scale. It is very much required that frame length of the signal should be kept optimized. If the frame size is much shorter then we don't have sufficient samples to get a reliable spectral estimation. On the other hand, if the frame size is longer than the signal changes too much throughout the signal. A pre emphasis filter of the form $H(z) = 1 - (0.97)z^{-1}$ is applied first. The FFT (Fast Fourier Transform) analysis is performed using Hanning windows of duration 20ms, with the 10ms shift between frames for a sampling frequency of 16 kHz. A band pass filter of 40 gamma tone of channel is considered in the filter bank. A band pass filter is used in auditory modeling to approximate the frequency for small selected portion. The linear predictive coefficient and the estimated energy are transformed to LPC cepstrum and logarithmic energy respectively. The LPC cepstrum coefficients are depended on the peak-weighted. The smoothed spectral envelop is obtained by Fourier transformation. The time functions of the LPC (linear predictive coding) cepstrum coefficients are called as cepstrum coefficient is used in the feature extraction. Therefore, the logarithmic spectral envelop, which corresponds to the dynamics emphasized cepstrum be obtained as follows.

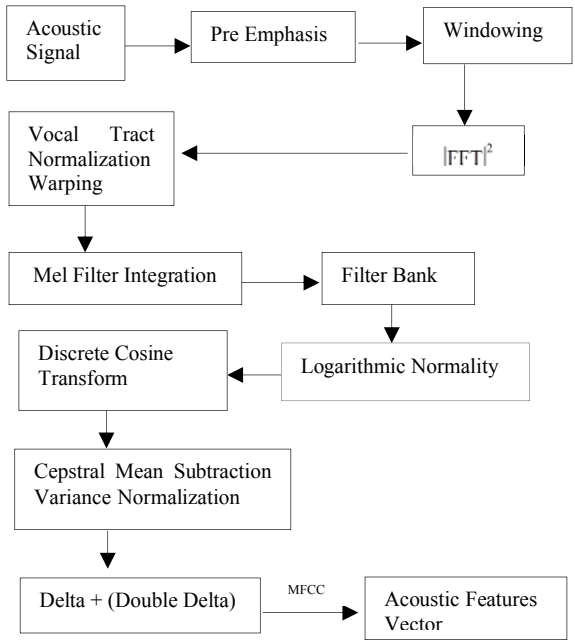

**Figure 1. Feature Extraction**

The $k$-dimension cepstrum vector, each of which consists of the 1$^{st}$ through the $k^{th}$ cepstrum coefficients for the adjacent $n$ – frames are denoted by $C_J$ for $(J = 1, \cdots, n)$.
The 1$^{st}$ and 2$^{nd}$ order polynomial expansion coefficients with their time functions are considered. If $S(w,t)$ and $C_{[m]}(t)$ are the power spectral envelop and the $m^{th}$ order cepstrum coefficient at time $t$ then it can be expressed as $\log(S(w,t)) = \sum_{-k}^{k} C_{[m]}(t) e^{-jmw}$. The data Cepstral features capture dynamic speech information that improves the speech recognition procedure. The Delta Cepstral and double-Delta Cepstral coefficients are appended to 13 MFCC features and 26 Delta- Cepstral coefficients. The Cepstral sequence is $C_{[m]}$ for short time interval. The Delta-Cepstral features are defined as,

$$\Delta_{[m]} = C_{[m+s]} - C_{[m-s]} \tag{1}$$

The index $m$ is related to considered frame for analysis and $s$ is an integer constant. The symbol $\Delta$ operator is stood for Delta operation. The double –Delta Cepstral features are obtained by the Delta operation on the expression (1) again. The acoustic feature vectors are obtained according to the procedure presented in figure 1.

## 3. DIMENSION COMPACTNESS OF TOTAL VARIABLITY SPACE

Joint factor Analysis (JFA) considers the $K$-number of acoustic feature vectors that is obtained according to the flow diagram 1. The JFA is implemented on the $m$ probabilistic pattern of the linear Gaussian mean supervector. Here $K$ is the number of observed feature vectors obtained according to the flow diagram 1, these $K$ vectors are independent and identically distributed.

Now, $Q \supset \{q_t\}_{t=1}^{K}$ with $q_t \in R^F$, $Q$ is the collection of all possible acoustic features and $F$ is the dimension of the acoustic class. The observed vector $q_t$ is presented by $L$-components soft Gaussian Mixture model (GMM), for the utterance model $\lambda = (\{\omega_i\}, \{\theta_i\}, \{\sum_i\})$.

The general $d$– variate GMM is expressed as,
$p_\lambda(q_t | \{\theta_i\}) =$
$\sum_{i=1}^{L} \omega_i \frac{1}{(2\pi)^{d/2} |\sum_i|^{1/2}} \exp\left\{\frac{1}{2}\left((q_t - \theta_i)' \sum_L^{-1} (q_t - \theta_i)\right)\right\}$

The means $\theta_i$ for $i = 1, 2, \cdots, L$ are random vector $\theta_i \in R^F$, with associated weight, $\omega_i \in R$ and $\sum_{i=1}^{L} \omega_i = 1$ with covariance is $\sum_i \in R^{F \times F}$ for $i = 1, 2, 3, ... L$.

The Universal Background Model (UBM), which is a big size GMM, is built with likelihood function, $p(Q|\lambda) = \sum_{j=1}^{J} \omega_j p_\lambda(q_t | \theta_i, \sum_j)$, $q_t$ is the acoustic vector at time $t$, and $\omega_j$ is the mixture weight for the $j^{th}$ mixture component. Here, $p_\lambda(q_t | \theta_i, \sum_j)$ is a Gaussian probability

function with mean $\theta_i$ and covariance matrix is $\sum_j$. Now, $J$ is the total number of Gaussian present in the in the mixture UBM [14] and optimized value of $J$ is the number $(1024 \times i)$, for $i = 1, 2, 3, ...$

To adopt the Gaussian means of UBM, we consider 'Maximum A Posterior' (MAP) method according to the flow diagram 2. The mean supervector is constructed by appending together with the means of each mixture component. It is expressed as $\theta = \{\theta'_1, \theta'_2, \cdots, \theta'_L\}' \in R^{F*L}$, the notation '/' is the transpose of vector and $E(\theta) \approx m$.

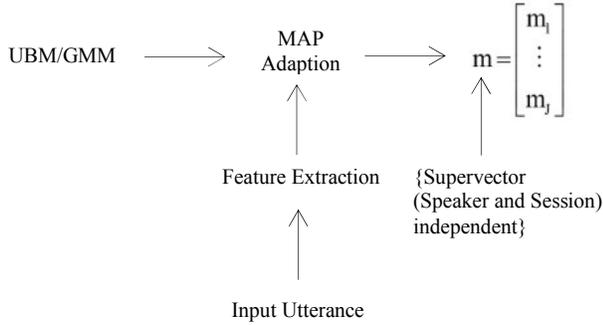

**Figure 2. Gaussian means Supervector**

The total variability modeling assumes the GMM mean super vector is $M$. It is optimized and presented by a set of feature vectors with the expression

$$M = m + \phi \quad (2)$$

Here, $m$ is the speaker and session independent super-vector from the universal background model (UBM), $\phi$ is an additive function of the speaker subspace and channel subspace. The JFA helps to reduce the subspace size. Prediction about the matching of input utterance based on "dimension compactness" of total variability space for the acoustic features are consider as the cosine value between the test utterance and target utterance. In this regards the support vector machine wouldn't much effective of enhancement for speech identification. The additive space is considered as a single space. This single space is considered as the total variability space. The variability space consists of the speaker variability and channel variability. The total variability space is presented by a total variability matrix $T(say)$. The total variability matrix is obtained from the collection of Eigen vectors for the corresponding larger Eigen values of the total variability covariance matrix of acoustic feature data-set. The data-set is obtained from the acoustic feature class data $Q$. The variability matrix $T$ is described as

$$T \begin{bmatrix} \sum X_1^2/N & \sum X_1 X_2/N & \cdots & \sum X_1 X_c/N \\ \sum X_2 X_1/N & \sum X_2^2/N & \cdots & \sum X_2 X_c/N \\ \vdots & \vdots & \vdots & \vdots \\ \sum X_c X_1/N & \sum X_c X_2/N & \cdots & \sum X_c^2/N \end{bmatrix}$$

Here, $\sum X_i^2/N$ is the variance for the $i^{th}$ component in the low dimensional total variability space and $\sum X_i X_j/N$ is the covariance of the $i^{th}$ and $j^{th}$ component of the total variability space, for $c, N \in I^{>0}$.

Now the expression (2) is modified to expression (3).

$$M = m + T\omega + \varepsilon \quad (3)$$

$T$ is the low dimension square matrix as it is obtained from the selected Eigan vectors of the corresponding larger Eigen values of the total variability space. The Eigan vector gives the direction along the maximum variability in low dimensional space. The matrix $T$ having the low ranks and $\omega$ is the random vector such that $\omega = \{\omega_i\}_{i=1}^c$, $\omega$ follows the standard normal distribution. The components associated with each vector $\omega_i$ are the feature factors collected from the matrix $T$. Those are collected according to the flow diagram 1. Each vector of $\{\omega_i\}_{i=1}^c$ follows identical type distribution, and hence it is called $'i-vector'$. $M$ is normally distributed with mean vector $m$ $\{m_1, \cdots m_J\}$, $M$ is session and channel dependent supervector with square covariance matrix $\sum TT'$. The '/' notation presents the transpose, and the notation $\varepsilon$ is the residual noise such that $\varepsilon$ is normally distributed with standard deviation $\sum$. The impact of noise is reduced on the total variability compact space. Hence, the equation (3) is redefined as

$$M \approx m + T\omega \quad (4)$$

## 4. PREDICTED COSINE SCORE MEASUREMENT

The $i-vector$ extraction is based on factor analysis. It is extended the session and speaker variabilities of super-vector to Joint Factor Analysis (JFA) [1, 13, 14]. The extracted $'i-vector'$ simultaneously and efficiently capture the speaker and channel variabilities. To identify any utterance in a target list, the cosine kernel based predicated procedure is implemented, between the test speaker $'i-vector'$ and target speaker $'i-vector'$. This procedure produces more optimized results. The cosine mapping based scoring procedure is effectively implemented as follows. Consider, two distinct speaker's utterances are $(x)$ and $(y)$ respectively. Those utterances are approximated by two distinct multinomial polynomials based on extracted acoustic features [7], say $P_x(i)$ and $P_y(i)$. Each multinomial polynomial consists of K number of features and each feature has at most F dimensions. The derived probabilities of feature are considered as $p_x(i\ 1)$,..., $p_x(i\ K)$ and $p_y(i\ 1)$,..., $p_y(i\ K)$ for two utterances. Since, $p_x(i)$ and $p_y(i)$ are the probability distributions for the corresponding utterances, clearly $\sum_{i=1}^{K} p_x(i) = \sum_{i=1}^{K} p_y(i)\ 1$. The divergence measurement between multinomial polynomials in the space $l(p)$ is defined a mapping $d: l(p) \times l(p) \rightarrow R^+$.

The mapping is expressed by

$$d(P_x, P_y)\ \cos^{-1}\left\{\left(\sum_i p_x(i) p_y(i)\right)^{q_k}\right\}^{1/S} \quad (5)$$

Let $\{p_k\}$ is bounded sequence of strictly positive real number and present the derived probability of features. For the two separate utterances we consider $k\ x, y$.

As, $\sup\{p_k\}\ 1$, since $q_k$ is the derived probability

So, $S\ \max(1, H)$ and $0 \prec p_k \leq \sup\{p_k\} = H$.

For simplicity, we consider $q_k\ \frac{1}{2}$ and $S\ 1$.

Now,

i) $d(P_x, P_x) = \cos^{-1}\left\{\left(\sum_i p_x(i) p_x(i)\right)^{1/2}\right\} = \cos^{-1}\{1\} = 0$

ii) $d(P_x, P_y) = 0 \Leftrightarrow x = y$

iii) $d(P_x, P_y)\ d(P_y, P_x)$

iv) According to figure 3.

We get $d(P_x, P_z) \prec (d(P_x, P_y) + d(P_y, P_z))$

According to figure 4.

We get $d(P_x, P_z) = (d(P_x, P_y) + d(P_y, P_z))$

The above two expression converges to

$d(P_x, P_z) \leq (d(P_x, P_y) + d(P_y, P_z))$

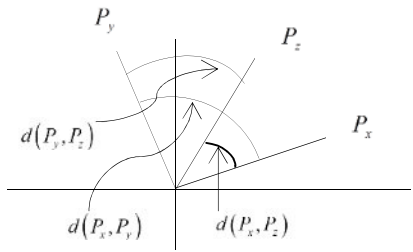

**Figure 3. Divergence measurement for 1st orientation**

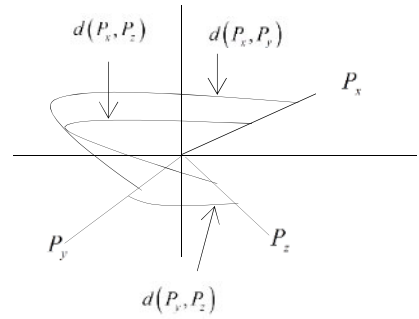

**Figure 4. Divergence measurement for 2nd orientation**

So, $l(p)$ is a metric space with metric $d$, (divergence measurement) that is expressed by the expression (5). According to the equation (4), the total variability modeling considers the GMM mean super vector is M. Those are the conversation side supervector, and it is clearly speaker and session dependent. The set of feature vector is decomposed according to expression (4). The mean supervectors from UBM are $\{m\}$ and those are speaker and session independent. The total variability matrix T spans a low dimensional subspace. Now, $\omega$ are the factors describing the utterance-dependent mean offset $T\omega$. The set of low dimensional total variability factors $\omega$ presents each conversation side. Each factor controls the separate Eigen dimension of the total variability matrix $(T)$. The distance scoring on channel compensated, 'i–vector' i.e. $\omega$ for a pair of conversation sides is the cosine between the target speaker, i-vector and the test speaker, i-vector in $l(p)$ metric space. The predicated score is obtained based on the cosine based prediction algorithm.

| Algorithm: Cosine based Prediction |
|---|
| **Input:** $w_{target}\ w_{test}$ <br> **Var** A: real <br> **Compute:** <br> $A\ \cos^{-1}\left(\frac{\langle w_{target}, w_{test}\rangle}{\|w_{target}\| * \|w_{test}\|}\right)$ <br> $if\left(0 \leq A \leq \frac{\pi}{2}\right) then$ <br> $score\ \cos(A);$ <br> $elseif\left(\frac{\pi}{2} \prec A \leq \frac{3\pi}{2}\right) then$ <br> $score\ 0;$ <br> $else$ <br> $score = \cos(2\pi - A);$ <br> **Output :** Score |

The acceptance or rejection based on user controlled decision threshold angle.

## 5. RESULT AND DISCUSSION

The above methodology being tested in the collected utterances from the noise full environment of telephone captured signal, mobile captured recording in rail station, busy bus stoppage from nonnative English spoke person. The corpus is the collected utterance from Native Indian languages likes Hindi, Bengali, Teague, and Oriya. In the testing purpose, we have randomly used English utterance as monologue or in the conversation even as a mixed spoke person. The used test utterance is of 45-second duration, and we consider the target list of 30 people. In this regards, the considered $i-vector$ with dimensions is 400 and feature dimension is equal to the 39 MFCC features. The tested results are presented in figure 5, figure 6 and figure 7. The figure 5 presents predicted cosine score about the probable matches for the speaker 1 speaker 2 and speaker 3. The horizontal axis presents the model identifier number that is already in the target list. The vertical axis presents the predicated cosine score between the test and target utterance in the range of [0, 1] with the scale 0.01. The score more than 0.9 to be considered as a good match with considering the false accept, and below 0.8 are not matched with considering false reject.

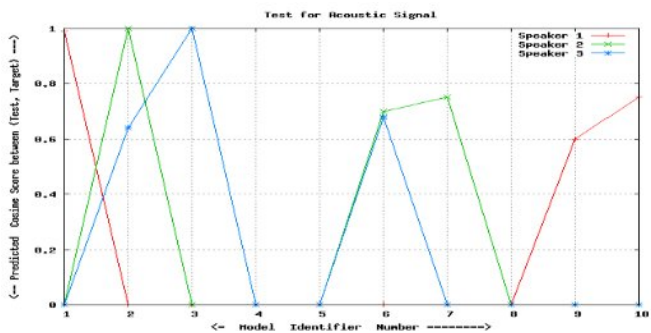

**Figure 5. Voice test score for speaker 1, 2, 3**

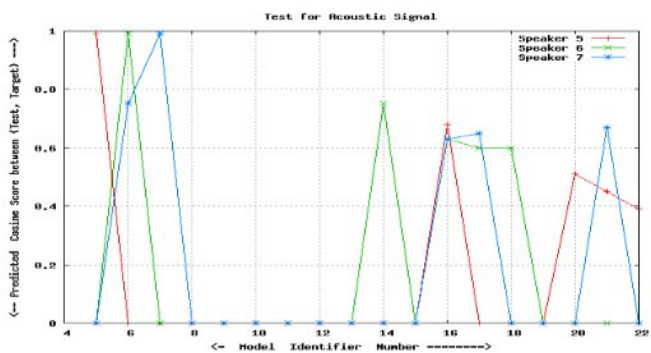

**Figure 6. Voice test score for speaker 5, 6, 7**

According to figure 7 if we consider 0.8 is the threshold level for acceptance, then the utterance of the speaker 9 that matches with the model identifier of the speaker 9. But the utterance of speaker-ID 9 is also matched to the model identifiers 11,25,26,27. Clearly, these 4 speaker models are falsely accepted. The equal error rates among the false accept and false reject controlled by the decision threshold. The decision threshold is user choice. It is depended upon the environment from where the utterance is collected.

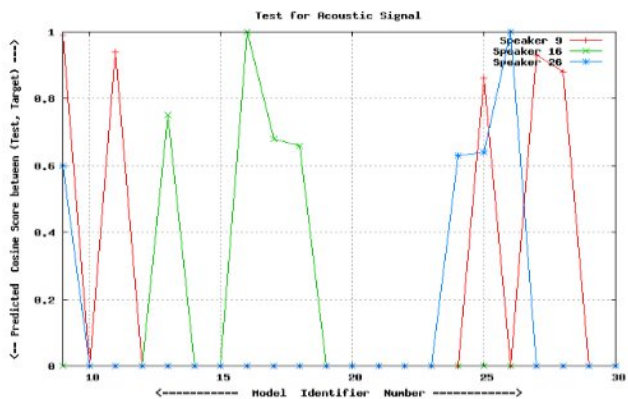

**Figure 7. Voice test score for speaker 9, 16, 26**

According to the figure 5, 6 and 7 if we consider the threshold is 0.9, then out of 30; only four speakers are falsely accepted i.e. 13%. If we consider a threshold is 0.8, then out of 30 speakers, only six speakers are falsely accepted i.e. 20%. If we consider the threshold is 1.0, then two speakers are falsely accepted i.e. 7%.

## 6. CONCLUDING REMARKS

This work presents the impact of dimension compactness in total variability space. The proposed methodology sufficiently reduces the computation time and works for small size of test utterance. The cosine scoring provides fast predicts about the matching. One of the most achievements is that if the predicted score is 1.0, it is the highly perfect match with little false acceptance by considering the noise full medium. The scoring particularly depends in which noise full environment the utterance being collected. The false accept and false reject depends on the decision threshold, but that could be controlled according to noise level of the environment. This methodology sufficiently reduces the false reject. For the specific suspicious target speaker, the proposed methodology enhanced the ASR system.

## 7. ACKNOWLEDGMENTS

Author would like to thank Niranjan Kumar Mandal for his continuous encouragement and motivation.